# FogGrid: Leveraging Fog Computing for Enhanced Smart Grid Network

Rabindra K. Barik
KIIT University,
Bhubaneswar, India
rabindra.mnnit@gmail.com

Satish Kumar Gudey
Gayatri Vidya Parishad College
of Engineering, AP, India
satishgudey13@gmail.com

Gujji Giridhar Reddy
Avanthi Engineering College
Visakhapatnam., AP, India
giri.gujju@gmail.com

Meenakshi Pant
KIIT University,
Bhubaneswar, India
meenakshikandpal14@gmail.com

Harishchandra Dubey
University of Texas at Dallas, USA
harishchandra.dubey@utdallas.edu

Kunal Mankodiya
University of Rhode Island, USA,
kunalm@uri.edu

Vinay Kumar
VNIT, Nagpur, India,
vinayrel01@gmail.com

*Abstract*—The present manuscript concentrates on the application of Fog computing to a Smart Grid Network that comprises of a Distribution Generation System known as a Microgrid. It addresses features and advantages of a smart grid. Two computational methods for on-demand processing based on shared information resources is discussed. Fog Computing acts as an additional layer of computational and/or communication nodes that offload the Cloud backend from multi-tasking while dealing with large amounts of data. Both Fog computing and Cloud computing hierarchical architecture is compared with respect to efficient utilization of resources. To alleviate the advantages of Fog computing, a Fog computing framework based on Intel Edison is proposed. The proposed architecture has been hardware implemented for a microgrid system. The results obtained show the efficacy of Fog Computing for smart grid network in terms of low power consumption, reduced storage requirement and overlay analysis capabilities.

## I. Introduction

Electric Power Utilities are guided by the SMART GRID (SG) to track and control the usage of power by the consumers. Its a real time bi-directional data communication performed by smart meters [1]. Load balancing in SG network is possible due to distributed energy management feature. The Utilities and the final consumers through online monitoring and controlling the power usage achieve the operation and management of their electric power systems. SG balances the energy consumption by sending the signals to the consumers whenever the energy consumption reaches peak values. It orders the consumers to cut off some of the unused appliances from the supply to overcome the peak demand [2]. In the control center, all the protection devices and the complete power system is well monitored for security purpose.

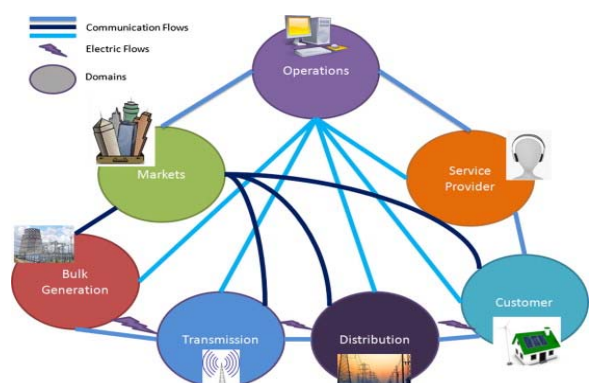

Fig. 1  Smart Grid conceptual model

Henceforth, an SG can be defined as an Electric grid using the information Technology as the communication barrier. For a layman, a SG consists of two parts namely the electrical power systems comprising the generation, transmission, distribution and utilization, the second consists of the information technology where data is to be transmitted and retrieved during necessity automatically [1]. It increases the efficiency, the reliability of power i.e. continuous supply of power and also the performance of the grid. The main issues in current electricity power markets are the reliability, efficiency, flexibility, adjustment of load, maximum power cut, market supply and demand response [3]. Demand side management helps in increasing the efficiency of the system. Load balancing is done by adjusting of load. Self-healing and fault detection abilities ensures reliability provided by the SG.

Fig. 1 shows the conceptual model of the SG given by the National Institute of Standards and Technology (NIST) [4], which elaborates the functions of a SG in terms of operations, requirements, characteristics, services that should be provided. It consists of seven models as shown in Fig. 1. They are production of power, transmitting the power and distributing the power. Apart from those, support to customers y providing them service during operations and educating them the present market scenario and the consumption. All these models help in many benefits including efficiency, low cost, fault tolerance and renewable energy generation.

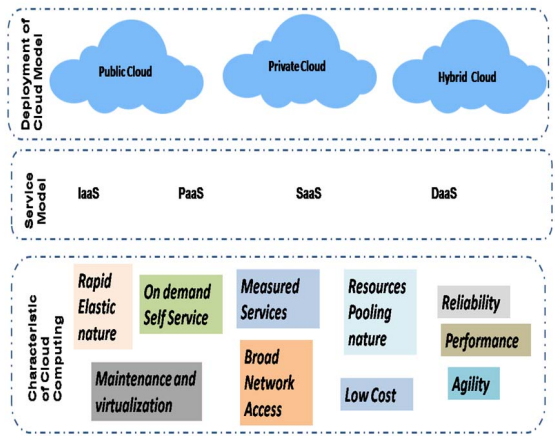

Fig. 2 Cloud computing applications.

SG's involve many intelligent meters and intelligent appliances, such as smart washing machines, Ovens, dishwashers etc. All these appliances are connected to the Communication Network i.e. the Internet [5][13]. Many sensors are required to be used in these devices to make sure that every moment be identified, maintained and regulated. Intelligent modules are to be used on the energy provider side to balance the upcoming demand and present supply. SG's tend to narrow from a Organized level to a regional level automatically by taking instantaneous decisions on its own. To make the grid more smarter, various technologies, applications are available, e.g., advanced metering infrastructure (AMI), Distribution Management System (DMS), Geographic Information System (GIS), Outage Management Systems(OMSs), Intelligent Electronics Devices (IEDs), Wide Area Measurement Systems (WAMS) and Energy Management Systems (EMSs) [6]. All the defined systems are These systems are guided effectually by Internet of Things (IoT) [7]. Using IoT machines and applications are connected well and understand the surroundings in a better way and make some intelligent decisions [5].

Both Cloud computing and Fog Computing are the advanced data processing models that contribute to on-demand competencies, and shared reserves on the Internet. Cloud computing depends on heavy memory and intelligent devices and act as a source provider [1], [2]. Fog based Computing on the other hand is a highly pragmatic platform that extends the cloud computing within the network framework so that the computation processes, storage data, and communication networking services can be done locally. Fog computing has caused a revolution in healthcare and remote care of patients, we are here translating those research to smart grid for societal benefit.

In this paper, Cloud Computing and Fog Computing based Smart Grid Network is discussed together with the Microgrid application of the SG network. Section II deals with the Cloud computing with its applications in SG, Section III discusses the Fog computing model, Section V the microgrid application using Fog computing and its decision making in a Distribution Generation and finally the conclusions with future scope of work in section VI.

Fog computing has caused a revolution in healthcare and remote care of patients, we are here translating those research to smart grid for societal benefit. Fig. 2 shows the details.

## II. CLOUD COMPUTING

There are four distinct types of services in cloud computing namely Platform as a Service (PaaS), Software as a Service (SaaS), Infrastructure as a Service (IaaS) and Database as a Services (DaaS) [3], [4], [5][6] [19]. These four services are shown in Fig. 3.

All these services are embedded in the middle layer between the cloud computing and deployment of cloud computing. model shown in Fig. 2. Each service provides a special feature which is inherent to itself. IaaS includes storage and virtual machines. All the applications within the cloud are supported by SaaS which can be accessible through Web browsers. The fourth service model DaaS provides a virtual database for the user to operate without the use of a hardware or a software equipment or a new configuration for performance. The main advantages of Cloud Computing are the elastic nature, shared architecture, metering architecture and internet services [7], [8].

In the conventional Cloud Computing Framework, the data to the cloud server is sent where they are further analyzed and processed and take high processing time along with high internet bandwidth. So to overcome this problem Fog computing is used by some authors [2]-[7], [8].Cloud computing can be used in a SG network however there are some associated drawbacks by SG utilities. SG requires immediate, instantaneous decisions regarding the real-time computing and storage capacity. Even though there are many issues pertaining to cloud computing, it is a simple and accurate method for studying the SG requirements. The SG offers many advantages to consumers, however there is always required an improvement to be done to handle more secure, data efficiently and expandable systems.

Electric power systems consist of three subsystems namely production, transfer of power and utilizing the power. In today's scenario, Cloud Computing has been a prospect technology that develops all these SG subsystems and hence an important component for all applications of SG. Many well known projects have been implemented using Cloud Computing such as Globus [8], EGI-InSPIRE (Integrated Sustainable Pan-Europe an Infrastructure for Researchers in Europe) [9], Information Power Grid from NASA [8], Open

Nebula [8] and T Clouds [8]. The architecture of Cloud Computing varies with the applications of SG network.

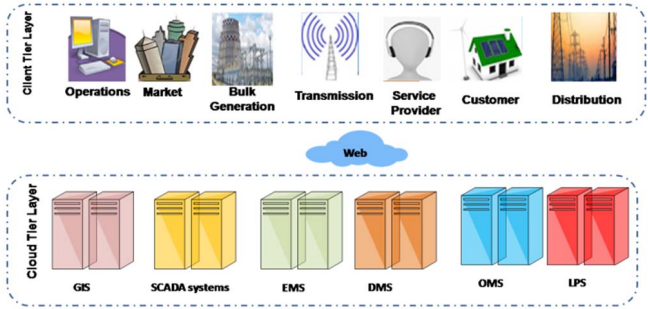

Fig. 3 Cloud Computing Framework of smart grid system.

For smart grid applications, Cloud computing is one of the most important technique. One of the best example is the integration of plug in hybrid Electric vehicles (PHEVs) using cloud computing to relieve the load demand from micro-grids. The customers have to pay more charges during the peak hours for charging their vehicles which increases the load on the SG [10], [11], [13]. This can be effectively addressed using the cloud computing storage system. The main advantages of CC over traditional models are energy saving, cost saving, agility, scalability, and flexibility, since computational resources are used on demand [12]. Fig. 4 shows the application of cloud computing with its service models discussed in this section.

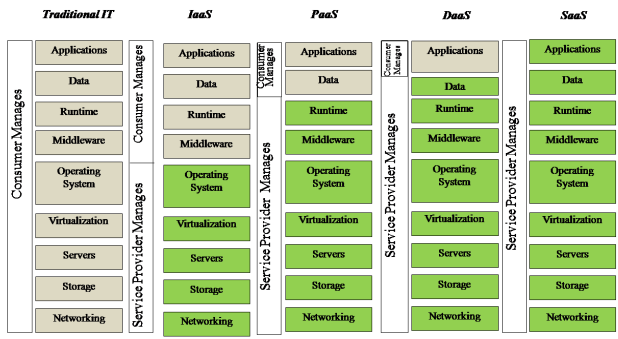

Fig. 4 Applications of the Cloud Computing service models.

### III. FOG COMPUTING

In Fog computing does not allow third parties to interfere in its network during communication with different smart devices and meters for processing and storage tasks [17] [18].Therefore, there is always a minimum need of the cloud operators.

Fig. 6 shows the proposed model of a fog computing architecture comprising of three tiers. The first tier consists of smart meters, the second tier consists of Fog servers, and the third one is the conventional cloud. Communication between the tiers is possible in four different ways. (a) smart device to smart device, (b) smart device to Fog server, (c) Fog server to Fog server, and (d) Fog server to cloud server.

Communication among the smart devices, smart appliances, electrical vehicles etc. is done by the first tier Com A, shown in Fig. 6, demonstrates the communication between different smart devices.

Take a case where an electric vehicle needs to be charged at a power outlet. The power outlet is not owned by the owner of the vehicle. It is located in the vicinity of the Fog server. For charging the vehicle, the smart meter of the outlet must be connected to the outlet of the owner which is in his/her house through communication. Hence a smart meter communication helps in charging the vehicle with the owner's ID and the energy consumption is cast back in the owner's bill.

Hence there is no need to move to a higher level in the proposed model. But if any two smart devices need to be connected which are not in the same fog area, then communication can be done in the higher level i.e. the fog tier as shown in Fig. 6. Take a case where the owner is riding his car in the fog server area and need to charge his vehicle. The procedure is to first the data is calculated in the smart grid tier and then is sent to the tier. The communication between the fog servers determines the data is transferred to the allotted fog server. Thereafter, the fog server gives the information to the car owner's smart meter using the identity of the vehicle [14]. This is shown as Com B in Fig. 5.

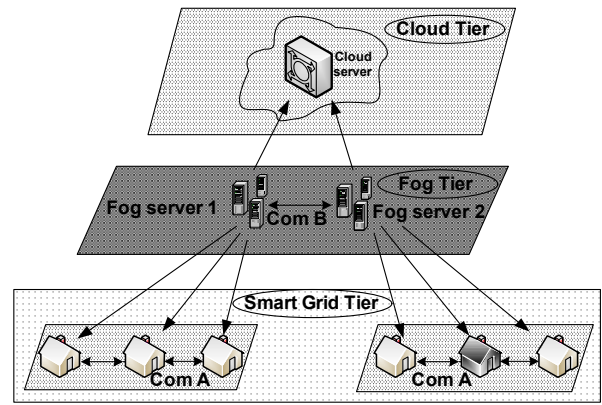

Fig. 5 Hierarchal communication structure of the Fog computing.

In a similar way, communication exists or takes place between different cloud servers if the smart meters are under different tiers. The fog tier basically connects and interacts with different smart meters in its vicinity and also collects data and shares the data from the consumers. It demarcates the private and public data so that the data storage and transmission is reduced. Always the private data is collected and stored as encrypted data and the public data as a non-encrypted data. The middle tier cannot decode the encrypted data as there is a confidentiality maintained between the cloud server and the consumer.

Both Cloud computing and Fog computing have specific meaning for a service range with in the cloud computing and client tiers which provide the mutual benefit to each other and interdependent services that leads to the greater storage capacity, control and communication possible anyplace within the specified range. It is not realized as an alternate source of communication to cloud computing especially in a smart grid network. In terms of locality, proximity, privacy, geo-distribution, location awareness and aggregation, Fog computing acts as a good resource of multi latency compared to cloud computing.

## IV. PROPOSED FRAMEWORK FOR SMART GRID APPLICATION

This section discusses distinct components of the Fog Computing Framework for smart grid applications and discusses the methods implemented in it. It uses Intel Edison as Fog computing device in proposed Framework. Intel Edison is powered by a rechargeable lithium battery and contains dual-core, dual-threaded 500MHz Intel Atom CPU along with a 100MHz Intel Quark microcontroller. It carries a memory of 1GB, 4GB RAM and supports IEEE 802.11 a, b, g, n standards. It connects to WIFI and has been used UbiLinux operating system for running compression utilities. Fig. 6 shows the proposed Fog computing Framework. The Fog device acts as a gateway between thick, thin and mobile clients and cloud layer.

The proposed Framework has three layers as client tier layer, Cloud computing layer and Fog computing layer. In client tier, the categories of users have been further divided into different categories. Processing of data can be possible within these three environments. Cloud computing Framework layer is mainly focused on overall storage and analysis of data. In Cloud computing Framework layer, The Fog computing layer works as middle tier between client tier layer and Cloud computing Framework layer. It has been experimentally validated that the Fog computing layer is characterized by low power consumption, reduced storage requirement and overlay analysis capabilities. In the Fog computing layer, the entire fog node has been developed with Intel Edison processor for processing of data. Large scale processing system (LPS) is the system in which we are using spark for big data processing.

## V. MICROGRID NETWORK USING FOG COMPUTING

A microgrid [16] is a part of the main grid working independently and in unison with the main grid. It has its own distribution electrical energy resources and loads. It operates in two modes namely the grid connected mode and the autonomous mode. It is built to ensure reliable, local affordable power to the critical loads such as military equipments, hospitals, data centers which require continuous power supply. Consumer EV, rooftop photovoltaic systems, residential-scale energy storage, and smart flexible appliances constitute the driving technologies in forming a residential microgrid [16].

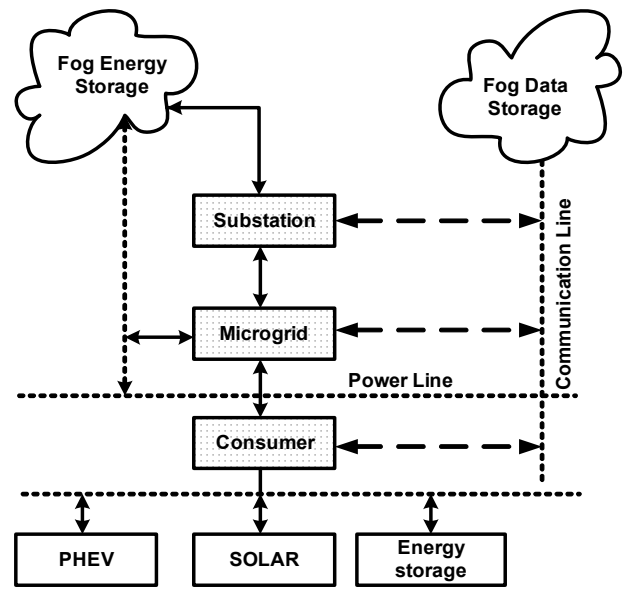

Fig. 7 Microgrid concept using Fog computing.

Fig. 7 shows the Mirogrid concept based on fog comptuing. The fog energy storage consists of the data of a BESS (Battery energy storage system) which may be a super conducting battery, used to store the energy during excess. It will be used to deliver power during critical conditions. The Fog data storage consists of the data pertainig to the substation, ,micro grid and the consumer. The comunication lines are used to transfer or retrieve the data from the data storage and sent to the cloud tier. The major sources of power are the solar in case of a micro grid represented here with an energy storage mechanism. The plug in hyrid electric vehicle is taken as an example to illustrate the mechanism of fog comptuing based smarg grid network.

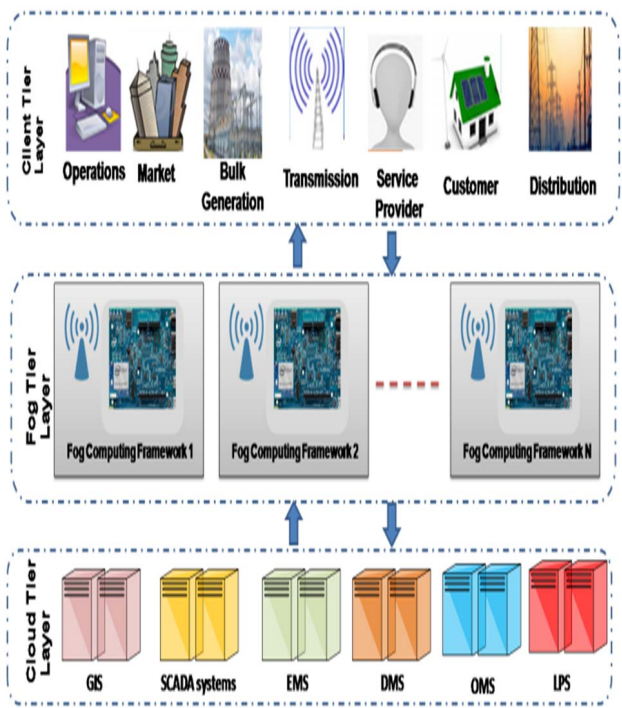

Fig. 6 Proposed Fog computing Framework.

As shown in Fig. 8 cloud computing is moved further to the edge of network (intelligence at the edge) and closer to the smart devices. It adds an intermediary layer to the platform providing the IoT with the ability of preprocessing the data while allowing low-latency requirements. It uses multiple on-the-edge (near user) devices to carry out maximum amount of communication, storage, control and processing capability. The developers can use their own security policy to send the data more securely as it is locally based compared to cloud computing. It also conserves the bandwidth and help the cloud to reduce the cost in transmitting huge data and infrastructural cost.

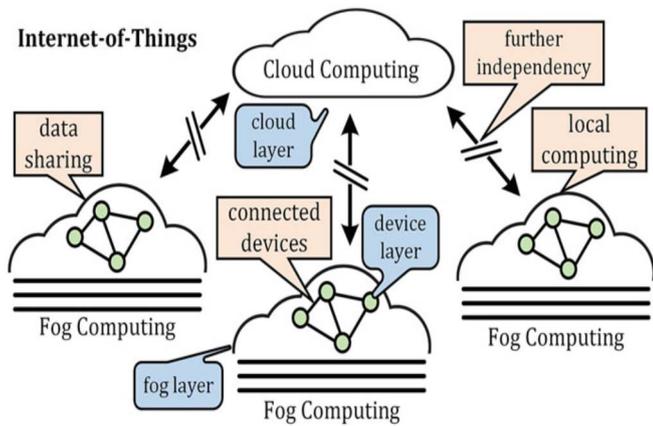

Fig. 8 Implementing Fog computing as an intermediary layer to cloud computing in Internet-of-Things (IoT) platform.

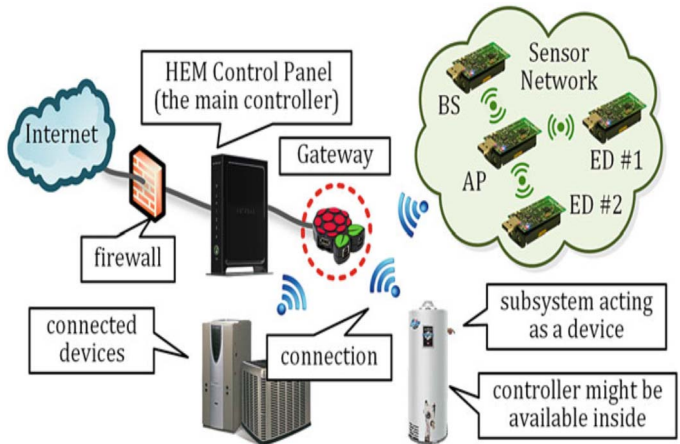

Fig. 9 Hardware architecture based on Fog computing platform for a Microgrid System.

The hardware architecture shown in Fig. 9 consists of multiple devices on an energy management platform like the home or a microgrid. These are classified based on their operational functionalities as connecting, gateway, sensor, actuator and computing.

## VI. RESULTS AND DISCUSSION

The microgrid network discussed in section V has been implemented in this section. The processing time and the average power consumption has been derived. In Fog Computing Framework, it has been used Intel edition processor in Fog node. Intel Edison produces processing time of order of NLog (N) where N defines the size of increased data sets. The main network has been designed in each Framework between the client tier layer and the Cloud computing Framework layer. It is assumed that the mean arrival rate of transmitted data would be once per minute assuming the Fog node can place in the locations where only a small number of devices in that area exist. The average waiting time for each and every Fog node has been calculated by Little's Law. It has been used which data for the different test of bench-marking experiment. It has also calculated the average load of the CPU, memory, processing time in percentage whereas average power consumption in watt. Fig. 10 shows the various performance comparisons between Cloud Computing Framework, and Fog computing Framework.

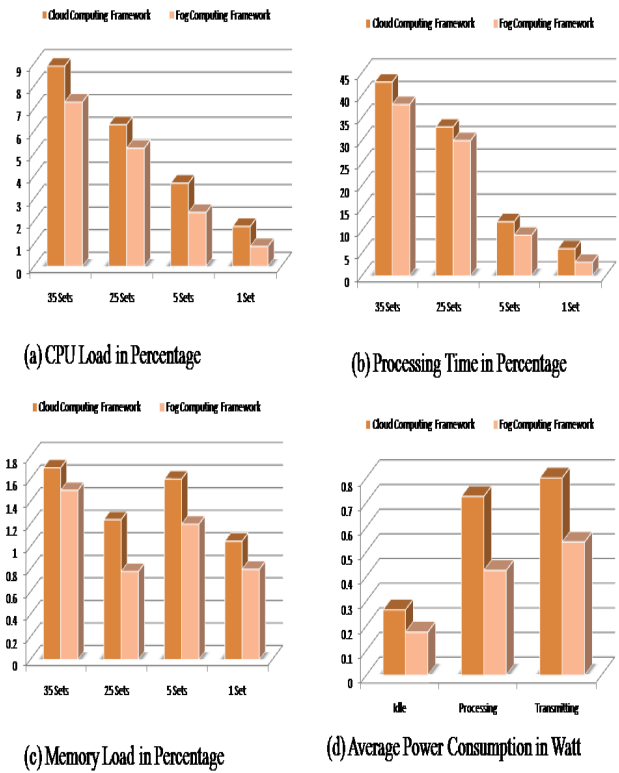

Fig 10 Performance comparison between Cloud computing Framework and Fog computing Framework.

From the comparison analysis, it shows that while running one set at a instant of time, the average waiting time for Cloud computing Framework is 188 seconds, the average waiting time for Fog Computing Framework is 84 seconds. It has also experimented that the service rate of Fog Computing

Framework is one third of Cloud Computing Framework. It is found that Fog Computing framework has been consuming 199mWs whereas cloud computing Framework has 489mWs when both these Frameworks are in active state.

## VII. CONCLUSIONS

In this paper, the application of fog computing for a microgid system is proposed, discussed and evaluated. The architecture of the fog and cloud computing are compared. It has been found that the fog tier acts as a secondary layer to cloud tire where the data can be stored and retrieved instantaneously and make the smart meters work independently. This type of architecture reduces the overall complexity in communication by taking immediate decisions that otherwise consume a lot of time. The main task is to educate the energy storage research personnel to upgrade such hierarchical structures for implementation. The future work rests in implementing different types of smart grids which are different in complexity and size. Due to the development of Internet of Things, smart grid can be implemented for all types of infrastructures.